\documentclass[10pt, conference, letterpaper]{IEEEtran}

\IEEEoverridecommandlockouts
\usepackage{cite}
\usepackage{amsmath,amssymb,amsthm, mathtools,amsfonts}
\usepackage{graphicx}
\usepackage{textcomp}
\usepackage{xcolor}
\usepackage{booktabs}
\usepackage{latexsym}
\usepackage{verbatim}
\usepackage{cite}
\usepackage[caption=false,font=footnotesize]{subfig}
\usepackage{subfig}
\usepackage{url}
\usepackage{epstopdf}
\usepackage{algcompatible}
\usepackage[noend]{algpseudocode}
\usepackage{times}
\usepackage{multirow}
\usepackage{enumerate}
\usepackage{xspace}
\usepackage{color}
\usepackage{mathrsfs}
\usepackage{mathtools}
\usepackage{epstopdf}
\usepackage{algorithm}
\usepackage{footnote}
\makesavenoteenv{tabular}
\algnewcommand{\algorithmicand}{\textbf{ and }}
\algnewcommand{\algorithmicor}{\textbf{ or }}
\algnewcommand{\OR}{\algorithmicor}
\algnewcommand{\AND}{\algorithmicand}
\algnewcommand{\var}{\texttt}

\newcommand{\dbmatrix}{\ensuremath {\boldsymbol{\mathcal{D}}}{\xspace}} 
\newcommand{\dbblock}{\ensuremath {\mathit{b}}{\xspace}} 
\newcommand{\dbrow}{\ensuremath {\mathit{r}}{\xspace}} 

\newcommand{\su}{\ensuremath {\mathit{SU}}{\xspace}}
\newcommand{\db}{\ensuremath {\mathit{DB}}{\xspace}}
\newcommand{\pu}{\ensuremath {\mathit{PU}}{\xspace}}
\newcommand{\pir}{\ensuremath {\mathit{PIR}}{\xspace}}
\newcommand{\sas}{\ensuremath {\mathit{SAS}}{\xspace}}
\newcommand{\epid}{\ensuremath {\textsc{EPID}}{\xspace}}

\newcommand{\rlist}{\ensuremath {\mathcal{L}}{\xspace}}
\newcommand{\protocol}{\ensuremath {\mathit{TrustSAS}}{\xspace}}
\newcommand{\misk}{\ensuremath {\mathcal{K}_{sk}}{\xspace}}
\newcommand{\mvpk}{\ensuremath {\mathcal{K}_{pk}}{\xspace}}
\newcommand{\umsk}{\ensuremath {\mathit{sk}}{\xspace}}
\newcommand{\epidSig}{\ensuremath {\mathit{\Sigma}}{\xspace}}
\newcommand{\cluster}{\ensuremath {\mathcal{C}{\xspace}}}
\newcommand{\beacon}{\ensuremath {\mathcal{\beta}{\xspace}}}
\newcommand{\beaconSet}{\ensuremath {\mathcal{B}{\xspace}}}
\newcommand{\epoch}{\ensuremath {\mathcal{T}_{epoch}{\xspace}}}
\newcommand{\threshold}{\ensuremath {t{\xspace}}}
\newcommand{\bc}{\ensuremath {\mathcal{BC}{\xspace}}}

\newcommand{\validators}{\ensuremath {\mathcal{V}{\xspace}}}
\newcommand{\tgdh}{\ensuremath {\textsc{TBLS}{\xspace}}}
\newcommand{\dkg}{\ensuremath {\textsc{DKG}{\xspace}}}

\newcommand{\beaconSUsSet}{\ensuremath {\mathcal{R}{\xspace}}}
\newcommand{\block}{\ensuremath {\mathcal{B}{\xspace}}}
\newcommand{\honestSet}{\mathcal{H}{\xspace}}
\newcommand{\dbsize}{\ensuremath {\eta {\xspace}}}
\mathchardef\mhyphen="2D

\newcommand{\algorithmicbreak}{\textbf{break}}

\newcommand{\algrule}[1][.2pt]{\par\vskip.5\baselineskip\hrule height #1\par\vskip.5\baselineskip}

\usepackage{soul}
\usepackage{color}
\makeatletter
\newcommand*{\whiten}[1]{\llap{\textcolor{white}{{\the\SOUL@token}}\hspace{#1pt}}}
\DeclareRobustCommand*\myul{%
    \def\SOUL@everyspace{\underline{\space}\kern\z@}%
    \def\SOUL@everytoken{%
     \setbox0=\hbox{\the\SOUL@token}%
     \ifdim\dp0>\z@
        \raisebox{\dp0}{\underline{\phantom{\the\SOUL@token}}}%
        \whiten{1}\whiten{0}%
        \whiten{-1}\whiten{-2}%
        \llap{\the\SOUL@token}%
     \else
        \underline{\the\SOUL@token}%
     \fi}%
\SOUL@}
\makeatother
\newcommand{\epidk}{\ensuremath {\textsc{EPID}}{\xspace}}
\newcommand{\tgdhk}{{{Cluster}}{\xspace}}

{\bfseries}{\rmfamily}
{\bfseries}{\rmfamily}
{\bfseries}{\rmfamily}
{\bfseries}{\rmfamily}
\newtheorem{mycorollary}{Corollary}{\bfseries}{\rmfamily}
{\bfseries}{\rmfamily}
{\bfseries}{\rmfamily}
{\bfseries}{\rmfamily}
{\bfseries}{\rmfamily}

\def\BibTeX{{\rm B\kern-.05em{\sc i\kern-.025em b}\kern-.08em
    T\kern-.1667em\lower.7ex\hbox{E}\kern-.125emX}}

\begin{document}
\title{{TrustSAS:} A Trustworthy Spectrum Access System for the 3.5 GHz CBRS Band\vspace{-0.2in}}
\author{Mohamed Grissa$^{\star}$, Attila A. Yavuz$^{\ddagger}$, and Bechir Hamdaoui$^{\star}$\\
$^{\star}$ \small Oregon State University, grissam,hamdaoui@oregonstate.edu\\
$^{\ddagger}$ \small University of South Florida, attilaayavuz@usf.edu 
\vspace{-0.05in}
}


\maketitle

\begin{abstract}

As part of its ongoing efforts to meet the increased spectrum demand, the Federal Communications Commission (FCC) has recently opened up $150$ MHz in the $3.5$ GHz band for shared wireless broadband use. Access and operations in this band, aka Citizens Broadband Radio Service (CBRS), will be managed by a dynamic spectrum access system (\sas) to enable seamless
spectrum sharing between secondary users (\su s) and incumbent users. Despite its benefits, \sas's~design requirements, as set by FCC, present privacy risks to \su s, merely because \su s are required to share sensitive operational information (e.g., location, identity, spectrum usage) with \sas~to be able to learn about spectrum availability in their vicinity.
In this paper, we propose \protocol, a trustworthy framework for \sas~that synergizes state-of-the-art cryptographic techniques with blockchain technology in an innovative way to address these privacy issues while complying with FCC's regulatory design requirements.

We analyze the security of our framework and evaluate its performance through analysis, simulation and experimentation. We show that \protocol~can offer high security guarantees with reasonable overhead, making it an ideal solution for addressing \su s' privacy issues in an operational \sas~environment.

\end{abstract}

\vspace{-2mm}
\section{Introduction}\label{inro}


\IEEEPARstart{T}he Federal Communications Commission (FCC) continues its effort towards promoting dynamic access to spectrum resources, and has recently promulgated the creation of the Citizens Broadband Radio Service (CBRS) in the 3.5 GHz band (3550 - 3700 MHz)~\cite{federal2015report}. This opens up previously protected spectrum, used by the US Navy and other DoD members, for dynamic and opportunistic spectrum sharing. In its CBRS report~\cite{federal2015report,federal2016report}, FCC prescribes the use of a centralized spectrum access system (\sas) to govern CBRS sharing among incumbent and secondary users.
Like the case of TV white space (TVWS) access, \sas~comprises multiple geolocation spectrum databases (\db s), operated (typically) by different administrators and are required to communicate amongst themselves to ensure frequency use information consistency. Also, like in TVWS, \su s need to query the \db s~using their exact location information to be able to learn about CBRS spectrum opportunities in their vicinity.

\sas~supports three types of users: primary users (\pu s), priority access license (PAL) users, and general authorized access (GAA) users.
\pu s are top/first tier users with the highest priority, while new CBRS users, considered as secondary users, operate either at the second tier as PAL users or at the third tier as GAA users~\cite{ye2016overview}. PAL users are assigned through competitive auction and have priority over GAA users, but they are required to vacate the spectrum upon the return of \pu s. GAA users, on the other hand, operate opportunistically, in that they need to query the \db s~to learn about which spectrum portions are available---not being used by higher tier (\pu~or PAL) users.
%
Even though both PAL and GAA users are considered as secondary users, in the remaining parts of this paper, for ease of illustration, \su~refers to a GAA user, since only GAA users need to query \db s to learn spectrum availability; PAL users acquire spectrum access via bidding.
\vspace{-2mm}
\subsection{Key $\boldsymbol{\sas}$ Requirements}
\label{sec:req}
As stipulated by FCC~\cite{federal2015report}, \sas's capabilities will exceed those of TVWS~\cite{chen2015protocol}, allowing a more dynamic, responsive and generally capable support of a diverse set of operational scenarios and heterogeneous networks~\cite{clark2018trading}.
While some of FCC's design requirements for \sas, such as the ability to authenticate users, hold users accountable for rule and policy violation, and to protect against unauthorized database access and tampering, are similar to TVWS systems, other requirements are only specific to \sas, which include~\cite{federal2016report}:




\noindent $\bullet$ {\bf Information gathering and retention}:
\su s must keep \sas~informed about their current operating parameters and channel usage information at all time, so that \sas~can maintain accurate and up-to-date frequency usage information. While this is mandatory in \sas, it is only optional in TVWS.


\noindent $\bullet$ {\bf Coexistence}: \sas~is required to coordinate the interactions among PAL and GAA users to ensure interference-free coexistence among all CBRS users~\cite{federal2016report,marshall2017three}. This is different from TVWS systems, which focus primarily on protecting \pu s, and not on ensuring coexistence among \su s.

%

\noindent $\bullet$ {\bf Auditability}: \sas~must maintain audit logs of all system operations~\cite{winnf2017cbrscomm}, including \db~write operations, user membership status changes, etc. \sas~uses these logs to verify and ensure compliance with regulatory rules and policies.

It is then important that these requirements be met when designing \sas. The challenge, however, is that meeting them gives rise to some serious privacy issues, thereby impacting the adoption of this promising technology.

\vspace{-2mm}
\subsection{Privacy Issues in $\boldsymbol{\sas}$}
A subtle privacy concern arises in \sas, which pertains merely to the fact that \su s are required to share sensitive operational information with \db s in order for them to obtain spectrum availability information~\cite{federal2016report}. This information, which may include \su s' sensitive data, such as their locations, identities, spectrum usage, and transmission parameters, may be collected by an adversary or a malicious \sas~administrator and be exploited for economic, political, or other purposes~\cite{winnf2016cbrsthreat}. For instance, fine-grained location information can easily reveal other personal information about \su s including their behavior, health condition, personal habits or beliefs~\cite{grissa2017locationsurvey}.

It may not be acceptable for most users to expose such a sensitive information, especially in the presence of malicious entities that can exploit it for malicious purposes~\cite{grissa2017locationsurvey,khalfi2018airmap,grissa2018unleashing}. Such privacy risks may hinder the wide adoption of this promising spectrum sharing technology. Calls are starting to arise within the wireless community to raise awareness about this issue as it is the case with Federated Wireless in their comments to FCC regarding its report and order~\cite{federal2016report}. Therefore, it is necessary to design mechanisms that can protect \su s' sensitive information while at the same time abiding by FCC's rules and policies prescribed for \sas.

\vspace{-2mm}
\subsection{Contributions and Paper Organization}
Most of \sas' rules require \su s to share a great deal of their sensitive information, which conflict with \su s' privacy objectives. As a result, we are facing a dilemma: On one hand, all \sas~entities need to comply with \sas's requirements to have a stable, interference-free radio environment. On the other hand, it is important to offer privacy guarantees to \su s so as to promote this new spectrum sharing technology.
This dilemma makes the task of designing \sas~mechanisms that provide privacy guarantees while complying with \sas's requirements and rules very challenging.

We strongly envision that the public's (long-term) acceptance of the \sas~paradigm will greatly depend on the robustness and trustworthiness of \sas~vis-a-vis of its ability to address these privacy concerns. Therefore, in this work, we propose \protocol, a trustworthy \sas~design framework that aims to achieve these two conflicting goals. More specifically, \protocol~combines and synergizes state-of-the art cryptographic techniques with blockchain technology in an innovative way to address these privacy issues while complying with FCC's regulatory design requirements.
To the best of our knowledge, this work is the first to address such issues within the context of \sas~and CBRS.








We first provide in Section~\ref{sec:sys_overview} a high-level overview of our framework to help grasp the big picture. Then, in Section~\ref{sec:protocol}, we provide a detailed description of the framework. The security analysis and performance evaluation are provided in Sections~\ref{sec:security} and~\ref{sec:perf}, and the paper is concluded in Section~\ref{sec:conclusion}.

\section{System and Framework Overview}
\label{sec:sys_overview}
In this section, we present the system architecture and provide a high-level overview of \protocol. Fig.~\ref{fig:sys} can be referred to throughout this section to facilitate the description.

\subsection{Architectural Components}\label{subsec:arch}
As illustrated in Fig.~\ref{fig:sys}, \protocol~comprises three main architectural entities: FCC, multiple \db s, and multiple \su s.
Without loss of generality, throughout the paper, we use FCC to refer to FCC itself, or to any trusted third-party entity that is appointed by FCC to act on its behalf.
In \protocol, FCC
is responsible for enforcing compliance with regulatory requirements, providing system keying materials, handling the registration of \su s, and granting them permissions to join \protocol. \protocol~leverages and relies on the existence of multiple \db s for spectrum access, each typically run by a different administrator. These \db s are assumed to be synchronized and to be sharing the same content, as mandated by FCC. Also, \protocol~supports multiple \su s, including a set of pre-registered \su s to be deployed specifically for playing the role of anchor nodes. These anchor \su s serve to establish a backbone peer-to-peer (p2p) network that can be discoverable and joinable by new \su s.


The content of each \db~can be viewed/modelled as an $\dbrow \times \dbblock$ matrix \dbmatrix~of size \dbsize~bits, where \dbrow~is the number of records in the database, each of size \dbblock~bits. Each record in \dbmatrix~is a unique combination of a cell number, representing the location, a channel number, and other transmission parameters (e.g., max transmit power, duration, etc). In \protocol, each record in \dbmatrix~contains a smart contract that is to be created by \db s to define channel usage rules, such as the maximum number of \su s allowed to transmit simultaneously in a given location, \su's maximum transmit power, etc. With these smart contracts, \protocol~ensures fair sharing of the spectrum resources, and limits the interference among \su s, thus satisfying the {\em coexistence} requirement, stated in Section~\ref{sec:req}.
For simplicity, we assume that channel usage is permitted over a fixed duration independently from the channel, and that \su s need to query \db s for an updated channel availability information periodically every \epoch, where \epoch~is a tunable system design parameter.
The geographical area serviced by \protocol~is modeled as a grid of $N\times N$ cells of equal sizes, and an \su's location is expressed through the grid's cell index.

\subsection{\protocol~Initial Setup}

  \begin{figure}[t]
 \centering
 \includegraphics[width=0.47\textwidth]{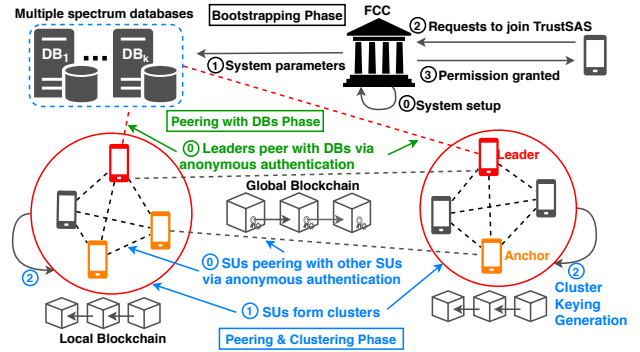}
 \vspace{-3mm}
 \caption{TrustSAS Architecture and Initial Operations}
 \label{fig:sys}
 \end{figure}

The first phase needed for setting up \protocol~is bootstrapping (see Fig.~\ref{fig:sys}), during which FCC creates the system parameters and keys, specific to \protocol, and shares them with \db s. Also, \su s first need to register and request \sas~access privileges from FCC before they can join \protocol. Once registered, FCC provides the joining \su~with the anchor \su~list, membership keys, and the procedure necessary for the \su~to authenticate with and join \protocol.
Note that, in \protocol, all messages communicated between the \su s and the \db s are established over secure channels, so as to ensure that the spectrum queries are authenticated, private, and not tampered with. Secure channels will be established via traditional mechanisms, and such mechanisms are ignored in this framework to keep the focus on the other security aspects. This phase is detailed in Section~\ref{subsubsec:boot}.

The second setup phase consists of establishing the underlying network infrastructure. Registered \su s that join \protocol~will maintain communication with one another via an overlay p2p network, and a newly joining \su~will rely on anchor \su s to discover and join the p2p network.\protocol~relies on an anonymous digital signature technique, explained in Section~\ref{subsec:setup}, to enable all these \su s to anonymously authenticate and verify each other's legitimacy when peering with one another. This anonymous authentication will also enable \su s to enjoy system services anonymously, yet in a verifiable way, to break the link between their sensitive operational data and their true identities.

\protocol~adopts a clustering approach, where joined \su s group themselves into clusters and elect cluster leaders, with the leaders being responsible for representing their \su s for interacting with other system entities. Not only will this improve \protocol~scalability, but also protect \su s' privacy, as it will limit the interaction with \db s to only cluster leaders.
Once clusters are established, \su s within each cluster distributively and collaboratively generate their cluster-specific keys, which will be used later for blockchain related operations inside the cluster and for signing cluster-wise spectrum agreements. This phase is detailed in Section~\ref{subsubsec:clustering}.

Once clusters are formed, the last setup phase is for the leaders to anonymously authenticate with \db s, and upon successful authentication, these \db s will join and be part of the established p2p network. This way, \db s will not be involved in the initial clustering of \su s, and therefore they will not be able to infer the \su s' location information. This phase is detailed in Section~\ref{subsubsec:peering-DBs}.

\subsection{\protocol~Main Operations}
%
\subsubsection{Querying Spectrum Availability Information} Each cluster leader acts on behalf of its \su~members and privately queries \db s for spectrum availability information. Even though the true identities of all \su s, including leaders, are hidden in \protocol, this is not sufficient to preserve their operational privacy. In fact, since each record in \db s is associated with a unique location, \db s may infer the location of the leaders from their queries and can still use this information for tracking purposes. To prevent this, \protocol~protects the leaders' queries through the adoption of {multi-server private information retrieval} (\pir) protocol~\cite{chor1995private}, which enables a user to retrieve a record from multiple databases while preventing the databases from learning any information about the record or the user requesting it. 
After learning the spectrum availability information, members of each cluster will distributively reach an agreement on how the spectrum resources are to be shared among them. Detailed description of this operation is provided in Section~\ref{subsec:query}.

\subsubsection{Notifying about Spectrum Usage}
Once a spectrum assignment agreement is reached, the cluster leader will notify the \db s about the spectrum portions used by its cluster members, as well as about other information, such as aggregate transmit power on each used channel, duration of channel use, etc., as required by FCC.
\protocol~uses this information to build knowledge of the spectral environment and to maintain an accurate availability information to comply with the information gathering and retention requirement. As we discuss in more details in Section~\ref{sec:protocol}, \protocol~ensures that cluster leaders report an accurate and non-altered spectrum usage information that is easily verifiable. Other leaders and \db s will distributively reach an agreement about the validity of this information, which, if valid, will be updated to \db s' records. Detailed description of this is provided in Section~\ref{subsec:notify}.

\section{The Proposed Framework: \protocol}
\label{sec:protocol}

%
%

\protocol~relies on permissioned blockchains~\cite{vukolic2017rethinking} to keep track of system and cluster activities. Blockchains are also used as a platform to handle agreements between entities at both the cluster and system levels. This is achieved thanks to permissioned blockchains' underlying {\em Byzantine fault tolerant} (BFT) consensus mechanism~\cite{vukolic2017rethinking}, which enables participants to reach agreements on block updates even when Byzantine nodes are present. Throughout the description of \protocol, before an entity submits and adds a block to a blockchain, we assume that the block is first signed by the entity and then validated via BFT by the validators of the blockchain. We now describe the different algorithmic components of \protocol.

\subsection{System Setup}\label{subsec:setup}
The first component of \protocol, depicted in Alg.~\ref{alg:setup}, consists of setting up the system parameters and the required keys at initialization, which is done in three phases.

\begin{algorithm}[b!]
\caption{{ \protocol~setup}}
\label{alg:setup}
\begin{algorithmic}[1]
\Statex\Function{TwoWayEPID}{$A,B,\mvpk,\rlist$}\label{step:twowayepid}
	\Statex User $A$ sends a challenge $m_A$ to user $B$
	\Statex User $B$ sends a challenge $m_B$ to user $A$
	\Statex $A$: ($\epidSig_A,\mathcal{P}_A) \gets$\epid.\textsc{Sign}($\umsk_A,\mvpk, m_B,\rlist$ )
	\Statex $B$: $v_A\gets$\epid.\textsc{Verify}($\mvpk, m_B,\epidSig_A,\mathcal{P}_A,\rlist$ )
	\Statex $B$: ($\epidSig_B,\mathcal{P}_B) \gets$\epid.\textsc{Sign}($\umsk_B,\mvpk, m_A, \rlist$ )
	\Statex $A$: $v_B\gets$\epid.\textsc{Verify}($\mvpk, m_A,\epidSig_B,\mathcal{P}_B,\rlist$ )		
	\Statex \Return $v_A \land v_B$
\EndFunction
\hspace{10pt}\algrule \vspace{-3pt}
\Statex  {\bf Bootstrapping phase}\vspace{-3pt}
\algrule
\State FCC: $(\mvpk,\misk) \gets \epid.\textsc{Setup}(\kappa)$ \Comment{$\kappa$: security level} \label{step:bootstrapping-begin}
\State FCC shares \mvpk~with \db s

\State  $(\umsk_\su,\mvpk)\mspace{-5mu} \gets$\epid.\textsc{Join(\mvpk,$\mathcal{K}_{FCC}$)} $\forall\su\mspace{-5mu}\in\mathcal{A}$

\For{$\su\:k\in\mathcal{A}$}
	\For{$\su\:l\in\mathcal{A}\setminus\{k\}$}
		\State $\textsc{TwoWayEPID}(k,l)$
	\EndFor
\EndFor

\State All \su s $\in\mathcal{A}$ peer up with each other
\State Joining \su: $(\umsk_\su,\mvpk) \gets$ \epid.\textsc{Join(\mvpk, $\mathcal{K}_{FCC}$)}
\State FCC shares $\mathcal{A}$ with joining \su \label{step:bootstrapping-end}
\hspace{10pt}\algrule \vspace{-3pt}
\Statex  {\bf Peering and clustering phase}\vspace{-3pt}
\algrule
\State \su~joins and discovers the p2p network through $\mathcal{A}$ \label{step:peering-clustering-begin}
\State \su~runs $\textsc{TwoWayEPID}()$ with each peer
\State \su s of the overlay network form clusters $\{\cluster^{(i)}\}_{1\leq i\leq n_\cluster}$
\State \su s $\in\cluster^{(i)}$ elect a leader $\su^{(i)}_L, \; \forall \;1\leq i\leq n_\cluster$
\State \su s $\in \cluster^{(i)}$ maintain a local blockchain $\bc^{(i)}$
\State \su s $\in \cluster^{(i)}$ run steps 2-6 of \textsc{Rekeying($\cluster^{(i)}$)} (Alg.~\ref{alg:rekeying}) \label{step:peering-clustering-end}
\hspace{10pt}\algrule \vspace{-3pt}
\Statex  {\bf Peering with \db s }\vspace{-3pt}
\algrule
\State \db s form validators set $\validators$ \label{step:peeringDBs-begin}
\State Global blockchain \bc~is created with validators $\in \validators$
\State \db s $\in \mathcal{V}$ and FCC maintain full copy of \bc
\For{$i=1,\cdots,n_\cluster$}
	\State $\su^{(i)}_L$ authenticates with \db s using \epid \label{step:epid}
	\State $\su^{(i)}_L$ peers up with \db s and becomes a validator
	\State $\su_L^{(i)}$ submits $y^{(i)}$ to \bc
	\State $\su_L^{(i)}$ requests a beacon $\beacon^{(i)}$ from a \db \label{step:beacon}
	\State \db~sends an \epid~challenge $m$ to $\su_L^{(i)}$
	\State $\su_L^{(i)}$:($\epidSig_L,\mathcal{P}_L)\mspace{-5mu}\gets$\epid.\textsc{Sign}($\umsk_L,\mvpk, m,\rlist$)
	\State \db~verifies ($\epidSig_L,\mathcal{P}_L)$ with \epid.\textsc{Verify}()
	\State \db~issues $\beacon^{(i)}$ to $\su_L^{(i)}$ and submits it to \bc
	\State $\su_L^{(i)}$ selects \su s $\in \cluster^{(i)}$ into $\beaconSUsSet^{(i)}$
	\State Every $\mathcal{T}_\beacon$, \su s $\in \beaconSUsSet^{(i)}$ transmit $\beacon_i$ for a duration $d$ \label{step:trans}
\EndFor
\end{algorithmic}
\end{algorithm}

\subsubsection{Bootstrapping Phase (Alg.~\ref{alg:setup}, steps~\ref{step:bootstrapping-begin}-\ref{step:bootstrapping-end})}\label{subsubsec:boot}
\protocol~ensures that \su s activities are anonymous, yet verifiable, by leveraging Intel's anonymous digital signature, known as enhanced privacy ID (\epid)~\cite{brickell2012enhanced}. \epid~allows any \su~to prove its membership legitimacy to other \protocol~entities, without revealing its true identity, using zero-knowledge proof~\cite{rackoff1991non}. \epid~also enables access revocation of misbehaving \su s anonymously, by maintaining and using a revocation list \rlist~based on \su s' signatures.
\epid~typically runs four procedures. The first, $\epid.\textsc{Setup}$,
is run by the FCC as the first step of the Bootstrapping phase (step~2, Alg.~\ref{alg:setup}) and outputs two system keys: {\em Membership Verification Public Key (\mvpk)} and {\em Membership Issuing Secret Key (\misk)}. The first key, \mvpk, is shared among all entities of \protocol, and used by \su s and \db s to anonymously verify the membership legitimacy of another \su. The second key, \misk, is kept secret and used only by FCC to create a unique {\em Membership Private Key}, $\umsk_{\su}$, for each joining \su, a key that the \su~uses to prove its membership legitimacy to the other system members anonymously. {\em We iterate again that FCC will be used throughout to refer to either FCC itself or any third-party entity that is appointed by FCC to govern on its behalf.}

The second procedure, $\epid.\textsc{Join}$, is run interactively between each joining \su~and FCC, and takes as input \mvpk~and FCC's public key $\mathcal{K}_{FCC}$, as illustrated in steps~4 and 9 of Alg.~\ref{alg:setup}. It results in \su~obtaining \mvpk~and $\umsk_\su$.
The third procedure, $\epid.\textsc{Sign}$, allows an \su~to anonymously prove its membership legitimacy and that it does not belong to the revocation list (i.e., its signature over a challenge message, $m$, does not belong to \rlist). Note that \epid~signatures produced by the same \su~are linkable; this prevents any malicious \su~from forging multiple signatures on behalf of other \su s. To validate the \epid~signature of joining \su, a verifier uses the fourth procedure, \epid.\textsc{Verify}, using the membership public key, \mvpk, by checking that \su's signature is not in \rlist.

\protocol~also requires that some \su s be appointed to serve as anchor nodes. These \su s need to run the \textsc{TwoWayEPID} subroutine (Alg.~\ref{alg:setup}, step~\ref{step:twowayepid}) among themselves to authenticate each other anonymously before they peer up and initiate the overlay p2p network. Later on, every joining \su, that obtained its $\umsk_{\su}$ through \epid.\textsc{Join}, will also get the list of anchor nodes, denoted by $\mathcal{A}$ throughout, from FCC.

\subsubsection{Joining and Clustering Phase (Alg.~\ref{alg:setup}, steps~\ref{step:peering-clustering-begin}-\ref{step:peering-clustering-end})} \label{subsubsec:clustering}
Every joining \su~uses the list $\mathcal{A}$ to discover and join the ongoing p2p network. The joining \su~then needs to authenticate with its peers and verify their legitimacy via \textsc{TwoWayEPID} (Alg.~\ref{alg:setup}, step~\ref{step:twowayepid}). After enough \su s have joined \protocol, these \su s will form clusters based on their locations; this may require the \su s to expose their locations to other \su s, but it should be no issue at this point since \db s are not part of the p2p network yet.
The members of each $\cluster^{(i)}$ will also maintain a cluster (local) blockchain, $\bc^{(i)}$, to log and keep track of key events taking place in the cluster.

\protocol~requires \su s of each cluster to serve as witnesses with respect to any cluster-related statement that is shared by the leader with the system. This is to prevent the leader from maliciously reporting incorrect information that was not validated by members of the cluster. To ensure this, \protocol~adopts the robust $(t,n)$-threshold BLS (\tgdh) signature scheme~\cite{boldyreva2003threshold}. \tgdh~requires no more than (any) $t_i+1$ of the $n_i$ \su s in $\cluster^{(i)}$ to collaboratively create a cluster signature over a statement. For this, members of each $\cluster^{(i)}$ will have to run the \textsc{Rekeying} operation described in Alg~\ref{alg:rekeying}, among the $n_i$ \su s of $\cluster^{(i)}$, to jointly generate the keys required for performing such distributed $(t_i,n_i)$-\tgdh~signatures within $\cluster^{(i)}$. This is achieved by running \tgdh's distributed key generation (\dkg)~\cite{gennaro1999secure} operation which will result in each $\su~j$ in $\cluster^{(i)}$ obtaining three keys: \tgdhk~Public Key, $y^{(i)}$, which is shared among all \su s in  $\cluster^{(i)}$, \tgdhk~User Secret Key, $x^{(i)}_j$, and \tgdhk~User Public Key, $z^{(i)}_j = g^{x^{(i)}_j}$. The \tgdhk~User Secret Keys $(x^{(i)}_1,\cdots,x^{(i)}_{n_i})$ are a $(t_i,n_i)$-threshold secret sharing of the private key $x^{(i)} = \log_g y^{(i)}$. These shares are constructed using Shamir secret sharing~\cite{shamir1979share} such that any subset of $t_i+1$ \su s from $\cluster^{(i)}$ can recover $x^{(i)}$ using Lagrange interpolation. \tgdhk~User Public Keys represent \su s' pseudonyms within $\cluster^{(i)}$ and are also used to identify \su s' transactions in the local blockchain, $\bc^{(i)}$. In addition to \dkg, \tgdh~comprises four other operations:
\begin{list}{\labelitemi}{\leftmargin=1em}
\item \textsc{SignShareGen}: It enables each \su~$j$ to compute the signature share $\sigma_j^{(i)}$ over a message $m$ to be signed by $\cluster^{(i)}$.
\item \textsc{SignShareVerif}: It enables members of $\cluster^{(i)}$ to verify \su~$j$'s  signature share $\sigma^{(i)}_j$ against its public key $z^{(i)}_j$.
\item \textsc{SignReconstruct}: The leader of a cluster collects a set of $t_i+1$ signature shares of a message $m$, $\mathcal{H}_i$, verified using \textsc{SignShareVerif}, from $t_i+1$ \su s. It combines these shares using Lagrange interpolation, via the Lagrange coefficients that were calculated in \textsc{DKG}, and reconstructs the complete cluster signature.
\item \textsc{GroupSignVerif}: Used to verify the cluster-generated signature against $\cluster^{(i)}$'s public key $y^{(i)}$.
\end{list}

Note that \tgdh~does not require reconstructing $x^{(i)}$ during the signing process. Even after repeated signing, no \su~could learn any information about $x^{(i)}$ that would enable it to create signatures without $t_i$ other \su s~\cite{goldfeder2017escrow}. We refer the reader to~\cite{boldyreva2003threshold} for more details about \tgdh.

%



\begin{algorithm}[h!]
\caption{Rekeying within $\cluster^{(i)}$}\label{alg:rekeying}
\begin{algorithmic}[1]
\Procedure{Rekeying}{$\cluster^{(i)}$}
	\State  $\{y^{(i)},x^{(i)}_1,\cdots,x^{(i)}_{n_i},z^{(i)}_1,\cdots,z^{(i)}_{n_i}\}\gets$$\tgdh.\textsc{DKG}(I)$
\For{$\su~j \in \cluster^{(i)}$}
\State ($\epidSig_j,\mathcal{P}_j)\mspace{-5mu}\gets$ \epid.\textsc{Sign}($\umsk_j,\mvpk, z^{(i)}_j,\rlist$ )
\State $\varrho_j\gets$ $\tgdh.\textsc{SignShareGen}(x^{(i)}_j,\epidSig_j,\mathcal{P}_j$)
\State \su~$j$ sends tuple $(\varrho_j,\epidSig_j,\mathcal{P}_j,z^{(i)}_j)$ to $\su^{(i)}_L$
\EndFor	
\State $\su^{(i)}_L$ submits $\{(\varrho_j,\epidSig_j,\mathcal{P}_j,z^{(i)}_j)\}_{j \in \cluster^{(i)}}$ to $\bc^{(i)}$	

	\State $\su_L^{(i)}$ submits $y^{(i)}$ to \bc
\EndProcedure

\end{algorithmic}
\end{algorithm}

To handle system-wise access revocations, \protocol~requires that each \su's \tgdhk~User Public Key is associated with its \epid~signature over some statement that is known by all cluster members. To achieve this, each \su~$j$ signs its \tgdhk~User Public Key $z_j^{(i)}$ itself, which is known to all \su s in the cluster, using \epid.\textsc{Sign} with its \epidk~Membership Private Key, $\umsk_{j}$ (Alg~\ref{alg:rekeying}, step 4). This serves to create a cryptographic binding between \su's \epid~signature and its \tgdhk~User Public Key. This binding will then have to be submitted as a transaction to be included in $\bc^{(i)}$. This is done by making \su~sign the binding from the previous step using \tgdh.\textsc{SignShareGen} with its \tgdhk~User Secret Key, $x^{(i)}_j$ (Alg~\ref{alg:rekeying}, step 5). Then each \su~will send the signatures, obtained in steps 4 and 5 of Alg~\ref{alg:rekeying}, to the leader $\su^{(i)}_L$, which will collect all these signatures and include them in $\bc^{(i)}$. 
Later, when an \su~$j$ is detected to be malicious, the leader will add \su's \tgdhk~User Public Key $z^{(i)}_j$ along with its \epid~signature to the revocation list \rlist.



\subsubsection{Peering with $\boldsymbol{\db s}$ Phase (Alg.~\ref{alg:setup}, steps~\ref{step:peeringDBs-begin}-\ref{step:trans})} \label{subsubsec:peering-DBs}
Each cluster leader will anonymously authenticate with \db s using \epid. Once a leader is authenticated by the \db s, these \db s join the established p2p network.


During this phase, a global blockchain \bc~is also created to keep track of the key system-wise events. Only \db s and cluster leaders can participate in the validation and addition of blocks to \bc. To submit a cluster-related block for inclusion in \bc, the leaders will need to have a key that identifies them and their clusters but also could be used to verify the correctness of the submitted block. This is exactly why each leader is required to submit its \tgdhk~Public Key, $y^{(i)}$, to \bc~to be shared with \db s and other leaders. On top of that, the leader will also share a $(t_i,n_i)$-\tgdh~signature of $y^{(i)}$ to show that the \tgdhk~Public Key was actually generated in collaboration with members of the cluster using \tgdh.\textsc{DKG}. The validators will validate the \tgdh~signature through a round of BFT consensus by verifying the signature against $y^{(i)}$.

In \protocol, an operational cluster is required to transmit a beacon for a certain duration, every $\mathcal{T}_\beacon$ period, so that the cluster could be discovered by nearby joining \su s, as in~\cite{chen2007cogmesh}. $\mathcal{T}_\beacon$ is a system design parameter that could be adjusted based on system dynamics and on how frequent \su s join the system. A leader $\su^{(i)}_L$ needs to request this beacon from one of the \db s and can acquire it only if it successfully proves its legitimacy to \db~through \epid~as depicted in steps~24-28 of Alg.\ref{alg:setup}. This is achieved by creating an \epid~signature of a challenge message $m$ that \db~has created for this purpose. If the \epid~signature is successfully verified, \db~issues a beacon to $\su^{(i)}_L$ and submits the beacon to \bc~so that it is accessible by all \protocol~entities. $\su_L^{(i)}$ picks some representatives from $\cluster^{(i)}$ to transmit the beacon every $\mathcal{T}_\beacon$, for a specific duration over a system control channel that is known a priori and is assumed to be reserved for this purpose.

Note that \su s in $\cluster^{(i)}$ only need to have a light copy of \bc~containing the latest state of the system including the current number of clusters and their corresponding beacons. Note also that a secure session is maintained between \db s and the leader of $\cluster^{(i)}$ as long as \epid~revocation list is not updated. This is to avoid running the \epid~verification protocol for every block or transaction submitted by $\su^{(i)}_L$.

\subsection{Joining \protocol}
As depicted in Alg.~\ref{alg:join}, when an \su~desires to join \protocol, it needs to tune to the control channel and scans it to detect any beacons transmitted by any nearby cluster. Failure to detect any beacons means that either no cluster is nearby or all nearby clusters are not accepting new \su s. In either case, \su~will start a new cluster and will request a beacon from one of the \db s and will itself start accepting new members, as described in Alg.~\ref{alg:setup}.

When the new \su~detects a beacon, it invokes the \textsc{TwoWayEPID} procedure with the cluster leader to ensure that the \su~is legitimate and can be allowed to join the cluster, and that the leader is also in a good standing. If the two-way verification is successful, the new \su~is admitted to the cluster and will immediately request $\bc^{(i)}$ from the cluster leader and peer with the \su s in the cluster. Newly admitted \su s will have to wait until the next $\epoch$ period to be able to participate in the cluster and enjoy spectrum resources.


Note that the admission of a new \su~to a cluster is also subject to interference constraints. Members of the cluster must ensure that the entry of this new \su~does not lead to an aggregate interference that is harmful to higher tier users or to other \su s in the cluster to satisfy coexistence. This could be resolved by adjusting grants and transmission parameters of the other \su s in the cluster, or simply denying the entry of a new \su~to the cluster in the extreme case. These scenarios could be enforced by the cluster leader and agreed upon through consensus among members of the cluster.

\begin{algorithm}[h!]
\caption{Join $\cluster^{(i)}$}
\label{alg:join}
\begin{algorithmic}[1]

\State $\su$~scans control channel for beacons in \beaconSet

\If{a beacon $\beacon^{(i)}$ of $\cluster^{(i)}$ is  found}
	\State $\su$~requests to join $\cluster^{(i)}$
	\State $v\gets$\textsc{TwoWayEPID($\su,\su_L^{(i)}$)}
	\If{$v == True$}
		\State $\su$~is added to $\cluster^{(i)}$
		\State $\su$~peers with \su s in $\cluster^{(i)}$ and downloads $\bc^{(i)}$
		\State $\su s \in \cluster^{(i)}$ run \textsc{Rekeying($\cluster^{(i)}$)} in next $\mathcal{T}_{epoch}$
	\EndIf
\Else
	\State \su~forms new $\cluster^{(i)}$ and becomes a leader $\su_L^{(i)}$
	\State $\su^{(i)}_L$ requests $\beacon^{(i)}$ as in Steps~\ref{step:beacon}-\ref{step:trans} of Alg.~\ref{alg:setup}

\EndIf

\end{algorithmic}
\end{algorithm}


Clusters will also need to perform rekeying operation when new \su s are added to their clusters, and this takes place at the end of each \epoch~period, where again \epoch~is a system design parameter that could be adjusted. Clusters could also choose to perform rekeying when malicious and/or faulty \su~s are detected. The rekeying steps are shown in Alg.~\ref{alg:rekeying}.




\subsection{Querying for Spectrum Availability}\label{subsec:query}
We now focus on describing the different steps required to privately query \db s for spectrum availability in a specific cluster. These steps are also summarized in Alg.~\ref{alg:pir}.

\begin{algorithm}[h!]
\caption{Private Spectrum Query}
 \label{alg:pir}
\begin{algorithmic}[1]

	\State $\su_L^{(i)}$ expresses interest to query \db s	\label{step:interest}	
	\State \db s send an \epid~challenge $m$ to $\su_L^{(i)}$
	\State $\su_L^{(i)}$: \epid.\textsc{Sign}($\umsk_L,\mvpk, m,\rlist$ )
	\State $\su_L^{(i)}$ requests other $\tau-1$ \su s to \epid.\textsc{Sign} $m$
	\State $\su_L^{(i)}$ sends $\tau$ \epid~signatures of $m$ to \db s
	\State \db s verify the $\tau$ signatures with \epid.\textsc{Verify}()	

	\If{any signature is not valid}
		\State \db~adds $\su_L^{(i)}$ to \rlist; \algorithmicbreak
	
	\EndIf
	
	\If{\su s $\in\cluster^{(i)}$ experience timeout from $\su^{(i)}_L$}
	\State \su s $\in\cluster^{(i)}\setminus\{\su^{(i)}_L\}$ elect new leader $\su^{(i)*}_L$
	\State $\su s \in \cluster^{(i)}\setminus\{\su^{(i)}_L\}$ run \textsc{Rekeying()}
	\State $\su^{(i)*}_L$ requests $\beacon^{(i)}$ as in steps~24-30 of Alg.~\ref{alg:setup}
	\State $\su^{(i)*}_L$ adds $\su^{(i)}_L$ to $\rlist$~and becomes $\su^{(i)}_L$
	\State go to Step~\ref{step:interest}
	\EndIf

	\State $\su_L^{(i)}$: $\dbmatrix_{\boldsymbol{q}} \gets$ $ \textsc{{BatchPIR}($\db s,\ell,t,r,s,{\boldsymbol{q}}$)} $\label{step:pir}
	\State $\su_L^{(i)}$ submits $\dbmatrix_{\boldsymbol{q}}$ as block $\block_{epoch}$ to $\bc^{(i)}$
	\State $\su s \in \cluster^{(i)}$ run BFT consensus to validate $\block_{epoch}$
	\State $\su_L^{(i)}$ triggers smart contracts to divide resources
	\State $\su s \in \cluster^{(i)}$ are assigned channels for current \epoch
\end{algorithmic}
\end{algorithm}

In \protocol, the cluster leaders will be in charge of querying \db s for spectrum availability on behalf of their \su~members, and a leader will query \db s only when: (i) Period allocated for using some channel(s) expires, (ii) quality of currently assigned channels degrades, or (iii) currently used channels need to be vacated (e.g., when requested by \pu s).

\subsubsection{Authentication and permission} In \protocol, in order for a leader to query \db s, its cluster is required to have a minimum of $\tau$ \su s, where $\tau$ is a system parameter that could be tuned depending on the desired robustness and security levels within each cluster.
Therefore, before querying the \db s, a cluster leader, $\su_L^{(i)}$, needs to show that its cluster $\cluster^{(i)}$ meets this requirement by providing $\tau$ \epid~signatures created by different legitimate \su s over a challenge message $m$ that \db s created for this purpose; this is depicted in steps 2-5 of Alg.~\ref{alg:pir}.
Note that \epid~prevents $\su_L^{(i)}$ from forging these $\tau$ signatures without being detected. Also, \protocol~will not require these $\tau$ \epid~signatures later unless a change in the membership of $\cluster^{(i)}$ takes place. If this verification is successful, then $\su_L^{(i)}$ proceeds with querying \db s for available channels. Otherwise, \db s will label $\su_L^{(i)}$ as malicious and add it to the revocation list, \rlist. To ensure robustness against a leader's failures, a timeout period could be considered beyond which if the \su~members do not receive spectrum availability information from their leader, the leader would be labeled as malicious and added to the revocation list, \rlist, and a new leader will be elected. The \textsc{Rekeying} procedure is then run among the cluster members, and the new leader will request a new beacon for the cluster as in steps~24-30 of Alg.\ref{alg:setup}.

\subsubsection{Spectrum querying} To enable private querying of \db s, \protocol~adopts multi-server private information retrieval (\pir) protocol~\cite{lueks2015sublinear}, termed \textsc{BatchPIR}, which leverages the multiple \db s, inherently available by \sas~design, to enable the cluster leaders to efficiently retrieve data records from \db s while preventing \db s from learning anything about the records being retrieved. It guarantees information-theoretic privacy, i.e. privacy against computationally unbounded servers, to cluster leaders as long as the spectrum database content, $\dbmatrix$, is replicated among  $\ell \geq 2$ non-colluding \db s~\cite{chor1995private}. The main idea consists of decomposing each leader's query into several sub-queries each processed by a different \db~to prevent leaking any information about the queried record.
BatchPIR also supports batching of the queries, i.e. retrieving multiple blocks simultaneously, which is a desirable feature for \protocol.
%
It takes as input the list of \db s, the maximum allowed number of colluding servers, the dimensions of \dbmatrix, and the indices of records of interest. For this, we assume that leaders can learn the index of the records of interest through an inverted index mechanism agreed upon with \db s.

A cluster leader, $\su_L^{(i)}$, collects queries from the \su s in its cluster $\cluster^{(i)}$, batches them together, and invokes \textsc{BatchPIR} with its peered \db s. $\su_L^{(i)}$ then submits the query response, $\dbmatrix_{\boldsymbol{q}}$, as a block $\block_{epoch}$ for inclusion in $\bc^{(i)}$. \su s in $\cluster^{(i)}$ run BFT consensus to validate this $\block_{epoch}$ by simply verifying the digitally signed database records against the public key of \db s. This is to prevent the leader from maliciously sharing altered availability information.

Each record in \db s contains a smart contract that defines its usage rules. Once $\block_{epoch}$ is validated by \su s and added to $\bc^{(i)}$, the scripts of the included smart contracts will reside in $\bc^{(i)}$. $\su_L^{(i)}$ will issue a transaction to trigger the execution of these smart contracts, which will take as input the list of \su s in the cluster, their cell indices, and the spectrum availability information. All this information is already stored in $\bc^{(i)}$ and is accessible by all \su s in $\cluster^{(i)}$. Once triggered, these smart contracts run independently and automatically in a prescribed and deterministic fashion on every \su's copy of $\bc^{(i)}$, in accordance with the data that was enclosed in the triggering transaction. The execution of these smart contracts will result in the automatic assignment of spectrum resources in a way that follows \protocol's guidelines while ensuring coexistence between \su s. This assignment will be valid for the duration of the $\epoch$ period. 
\vspace{-3mm}

\subsection{Notifying about Spectrum Usage}\label{subsec:notify}
\begin{algorithm}[h!]
\caption{Spectrum Usage Notification}\label{alg:usage}
\begin{algorithmic}[1]
\State $\su_L^{(i)}$ constructs block $\block_i$ with usage information
\State $\su_L^{(i)}$ sends $\block_i$ to \su s in $\cluster^{(i)}$ for validation and signing
\State $(\block_i,\sigma_{\block_i})\mspace{-5mu}\gets$\tgdh.\textsc{SignReconstruct}($\honestSet_i,L_1,\cdots,L_n$)
\State $\su_L^{(i)}$ submits $(\block_i,\sigma_{\block_i})$ to \bc
\State \validators:$val\mspace{-7mu}\gets\mspace{-7mu}\tgdh.\textsc{GroupSignVerif}(\block_i,\sigma_{\block_i},y^{(i)})$ $\!\!$ w/ BFT
\If{$val == True$}
	\State $\block_i$ is added to \bc
	\State \db s update their records
\Else
	\State \db s flag $\su_L^{(i)}$ as malicious
	\State $\su_L^{(i)}$ is added to revocation list \rlist~in \bc
	\State \db s remove $\beacon^{(i)}$ from list of valid beacons on \bc
\EndIf
\end{algorithmic}
\end{algorithm}

Once spectrum resources are allocated among \su s, the leader $\su_L^{(i)}$ shares with the \db s the allocation information, including the channels to be used by the members of $\cluster^{(i)}$, the locations where these channels will be used, and aggregated transmit power over those chosen channels. The leader can also collect the received signal strengths in the used and adjacent frequencies, the received packet error rates, and other standard interference metrics for all \su s in the cluster. The leader will propose a block $\block_i$ containing this information to the members of the cluster for validation. They will verify the correctness of this information and sign the block using \tgdh. If the validators successfully verify that $\block_i$ was agreed upon and signed by members of $\cluster^{(i)}$ via BFT consensus combined with \tgdh, then $\block_i$ is added to \bc~and \db s will include this information in their records. Otherwise, $\su_L^{(i)}$ will be flagged as malicious and its \epid~signature of $y^{(i)}$ will be added to \rlist. These steps are summarized in Alg.~\ref{alg:usage}. 
\vspace{-2mm}

\section{Security Analysis}
\label{sec:security}
\subsubsection{Threat Model}
\protocol~assumes that \db s are {\em honest-but-curious}, in that they act ``honestly'' and follow the protocol in terms of handling queries and sharing spectrum availability information, but they are also ``curious'' about \su s' information and try to infer it from the messages they receive from \su s.
\protocol~also assumes that these \db s do not collude with each other, nor with the \su s.
We refer to a \su~that faithfully follows the protocol as {\em honest}; otherwise, it is referred to as {\em Byzantine}. 
\protocol~assumes that these Byzantine \su s do not collude with \db s, and for each cluster $\cluster^{(i)}$, at least $t_i$~of the $n_i$ \su s participate in the signature, and no more than $f_i = (n_i-t_i)$ \su s are Byzantine.
%
These $t_i$ \su s serve as witnesses for the cluster to make sure that the leader does not communicate compromised information.

\subsubsection{Security Objectives}
Given the above threat model, \protocol~aims to achieve the following security objectives:

\noindent $\bullet$ \myul{\em Private Spectrum Availability Querying:} \su s can query \db s {\em privately}, without revealing their operational information.

\noindent $\bullet$ \myul{\em Private Spectrum Usage Notification:} \su s can notify \db s about their channel usage and transmission parameters {\em privately}, without revealing their operational information.


\noindent $\bullet$ \myul{\em Robustness to Failures:} All security guarantees are maintained, even when a system entity fails or is compromised.

\noindent $\bullet$ \myul{\em Immutable Public Log for Auditability:} A globally consistent, tamper-resistant log is maintained, where each system event, once produced and logged, cannot be altered or deleted.

\noindent $\bullet$ \myul{\em Anonymity and Membership Verifiability:} \su s' authenticity can be verified before the \su s are granted system access, and \su s cannot be identified at any stage of protocol execution.


\noindent $\bullet$ \myul{\em Location Privacy Protection of \su s:} \su s' physical location information is kept private at all times from all \db s.

\subsubsection{Security Results}
All proofs of the security results stated in this section are omitted here due to space limitation, and can be provided if and when requested.
\begin{mycorollary}
\label{cor:Unforg-Robust}
\protocol~achieves unforgeability and robustness of \tgdh~signatures against an adversary that can corrupt any $f_i < n^{(i)}/2$ \su s within a cluster $\cluster^{(i)}$ as long as the Gap-Diffie-Hellman problem is intractable.
\end{mycorollary}
\vspace{-0.1in}
\begin{mycorollary}
\label{cor:Consist-Resist}
\protocol~ensures consistency and resistance to fork attacks for a permissioned blockchain $\bc^{(i)}$ running BFT consensus in every $\cluster^{(i)}$ if $t_i\geq 2f_i+1$, where $t_i$ is the number of signature shares required to construct a group signature for $\cluster^{(i)}$, and $f_i$ satisfies $n_i \geq 3f_i+1$ for BFT mechanisms ~\cite{castro1999practical}.
\end{mycorollary}
\vspace{-0.1in}
\begin{mycorollary}
\protocol~guarantees unforgeability and robustness of \tgdh~signatures while ensuring consistency and resistance to fork attacks for $\bc^{(i)}$ of $\cluster^{(i)}$ against an adversary that can corrupt any $f_i < n^{(i)}/3$.
\end{mycorollary}
\vspace{-0.1in}
\begin{mycorollary}
\protocol~guarantees \su s with information-theoretic, private spectrum availability querying from \db s.
\end{mycorollary}
\vspace{-0.1in}
\begin{mycorollary}
\protocol~guarantees anonymous membership verification through \epid~as long as the Decisional Diffie-Hellman and the strong RSA assumptions hold and the underlying primitives they use are secure.
\end{mycorollary}
\vspace{-0.1in}
\begin{mycorollary}
\protocol~is robust against Byzantine failures of both \db s and \su s alike.
\end{mycorollary}
\vspace{-0.1in}
\begin{mycorollary}
\protocol~guarantees location privacy information protection to all \su s.
\end{mycorollary}


\section{Performance Evaluation}
\label{sec:perf}
We assess the effectiveness of \protocol~by evaluating the performance of its building blocks and algorithms. These evaluations are performed both analytically and empirically via either implementations or benchmarking of the underlying math and crypto operations using MIRACL library~\cite{miracl}.
Experiments are carried out on a testbed that we built on Geni platform~\cite{geni} using percy++ library~\cite{percy}. The testbed consists of $7$ VMs deployed on different Geni sites, each playing the role of a \db, and a Lenovo Yoga 3 Pro laptop with 8 GB RAM running Ubuntu 16.10 with an Intel Core m Processor 5Y70 CPU 1.10 GHz to play the role of a cluster leader.



\subsubsection{Distributed Key Generation (\dkg)}
Running \dkg~requires performing a number of elliptic curve point multiplications that is proportional to the number of \su s within the cluster. Using the benchmarking results that we derived with the MIRACL library~\cite{miracl}, we provide in Table~\ref{tab:tgdh} an estimate of the average processing time experienced by each \su~when running \dkg. In terms of communication overhead, \dkg~requires $2$ rounds of broadcasts, yielding $\mathcal{O}(n_i)$ messages per \su, or $\mathcal{O}(n_i^2)$ messages per cluster $\cluster^{(i)}$, when assuming no faulty \su s. Despite its relatively high cost, \dkg~presents no bottleneck to the system, as it is only executed at initialization or when group membership changes occur.

\begin{table}[h!]
\vspace{-0.1in}
\centering
\caption{\tgdh~Overhead within cluster $\cluster^{(i)}$}\label{tab:tgdh}
\vspace{-2mm}
\renewcommand{\arraystretch}{1.1}{
\begin{tabular}{lll}
\toprule
\textbf{Operation} \hspace{25pt} 				& {\textbf{Analytic Cost}}  \hspace{25pt} 	& {\textbf{Empirical Cost}} 			  \\
\midrule
\dkg~Computation  			&  $\mathcal{O}(n_i) \cdot PM$  & $1.05\; s$   \\
\dkg~Communication 					&  $\mathcal{O}(n_i)$ messages				& $\propto1000$ messages 		\\

\midrule

 \textsc{SignShareGen} & $1 \;Hash + 1\; Expp$ & $0.63 \;ms$ \\

  \textsc{SignShareVerif} & $2t_i\cdot TPO$ & $2.3\; ms$ \\
  Signature size & $64$ bytes & $64$ bytes \\
  Private key size & $32$ bytes  & $32$ bytes  \\
  \textsc{SignReconstruct} &  $t_i\cdot(Mulpp + Expp)$& $461\; ms$ \\
  \textsc{GroupSignVerif} &  $2\cdot TPO$& $2.3 \;ms$ \\
\bottomrule
\end{tabular}}
\flushleft{ \scriptsize{\textbf{Variables:} $PM$: cost of an elliptic curve point multiplication. $n_i = 1000$ \su s, $t_i = 1000$ \su s. $TPO$ is the cost of one tate pairing. $Expp$ and $Mulpp$ are the cost of a modular exponentiation and multiplication, respectively, over modulus $p$.

}}

\end{table}

\subsubsection{Threshold Signature (\tgdh)}
Table~\ref{tab:tgdh} provides the analytical and empirical cost of the different \tgdh~operations executed by \su s in $\cluster^{(i)}$.
\su s repeatedly sign the consensus statement at each BFT round within the cluster.
From an \su's perspective, this is relatively fast, as it involves signing a single message whose cost is dominated by a modular exponentiation operation, as shown in Table~\ref{tab:tgdh}. The leader, $\su_L^{(i)}$, will, however, incur most of the overhead, as it needs to verify all the signature shares coming from the $\threshold_i$ signing \su s of $\cluster^{(i)}$, before multiplying them to construct $\cluster^{(i)}$'s signature. These are the most expensive operations involved in \tgdh~as they require a number of modular multiplications and exponentiations that is linear in $\threshold_i$ as illustrated in Table~\ref{tab:tgdh}. To estimate the running time of \tgdh's different operations, we use dfinity's implementation of \tgdh~\cite{tblsdfinity}.

%
%
%
%

\subsubsection{Enhanced Privacy ID (\epid)}
We evaluate $\epid.\textsc{Sign}$ and $\epid.\textsc{Verify}$ analytically and empirically (using Intel's publicly available SDK~\cite{epidsdk}) as depicted in Table~\ref{tab:epid}. $\epid.\textsc{Sign}$ and $\epid.\textsc{Verify}$ both require a number of modular exponentiations that is linear in the size of the revocations sublists; these revocation sublists are defined in~\cite{brickell2012enhanced}.
 



\begin{table}[h!]
\vspace{-0.1in}
\centering  \caption{ \epid~complexity} \label{tab:epid}
\vspace{-2mm}
\renewcommand{\arraystretch}{1.1}{
\begin{tabular}{lll}
\toprule
{\textbf{Operation}}  & {\textbf{Analytical Cost}} &  {\textbf{Empirical Cost}}  \\
\midrule
$\epid.\textsc{Sign}$ & $(6\delta_2+2\delta_3+c) \cdot Expp$ &  $135 \;ms$\\
 $\epid.\textsc{Verify}$ &  $(\delta_1+6\delta_2+2\delta_3+c) \cdot Expp$& $120\;ms$ \\
 Signature size & $257$ bytes & $257$ bytes \\
 Private key size & $129$ bytes  & $129$ bytes \\
\bottomrule
\end{tabular}}
\flushleft{ \scriptsize{\textbf{Variables:} $\delta_i = \vert\rlist_i\vert$ for the revocation sublists~\cite{brickell2012enhanced} $\rlist_1$ (private key-based list), $\rlist_2$ (signature-based list), $\rlist_3$ (issuer-based list) with $ \rlist_1 \cup \rlist_2 \cup \rlist_3=\rlist$, and  $c$ is a constant. Cryptographic parameters correspond to 128-bit security level as in \cite{keylength}.
}}
\end{table}

Even though these delays seem relatively high, they are still reasonable, especially that these membership proof operations are independent, unfrequent, and do not occur simultaneously, once the system setup completes. Note that this proof has a linear cost in the size of the revocation list and could become quite expensive for both signers and verifiers if such a list becomes large. One possible way to maintain a good performance of \protocol~is to impose a threshold on the list size. In this case, when the list size exceeds the threshold, FCC can create a new group and perform a rekeying operation, with each \su~needing to prove to FCC that it is a legitimate member and that its membership was not revoked. This would be more efficient than carrying a large revocation list indefinitely and run expensive zero-knowledge proof operations on it. The old list will still be accessible for auditing purposes as it would have been stored already in \bc.

\subsubsection{Private Information Retrieval (\pir)}
\begin{figure}[!t]
\vspace{-0.1in}
\centering
\subfloat[\db~\pir~delay.]{\includegraphics[width=0.162\textwidth]{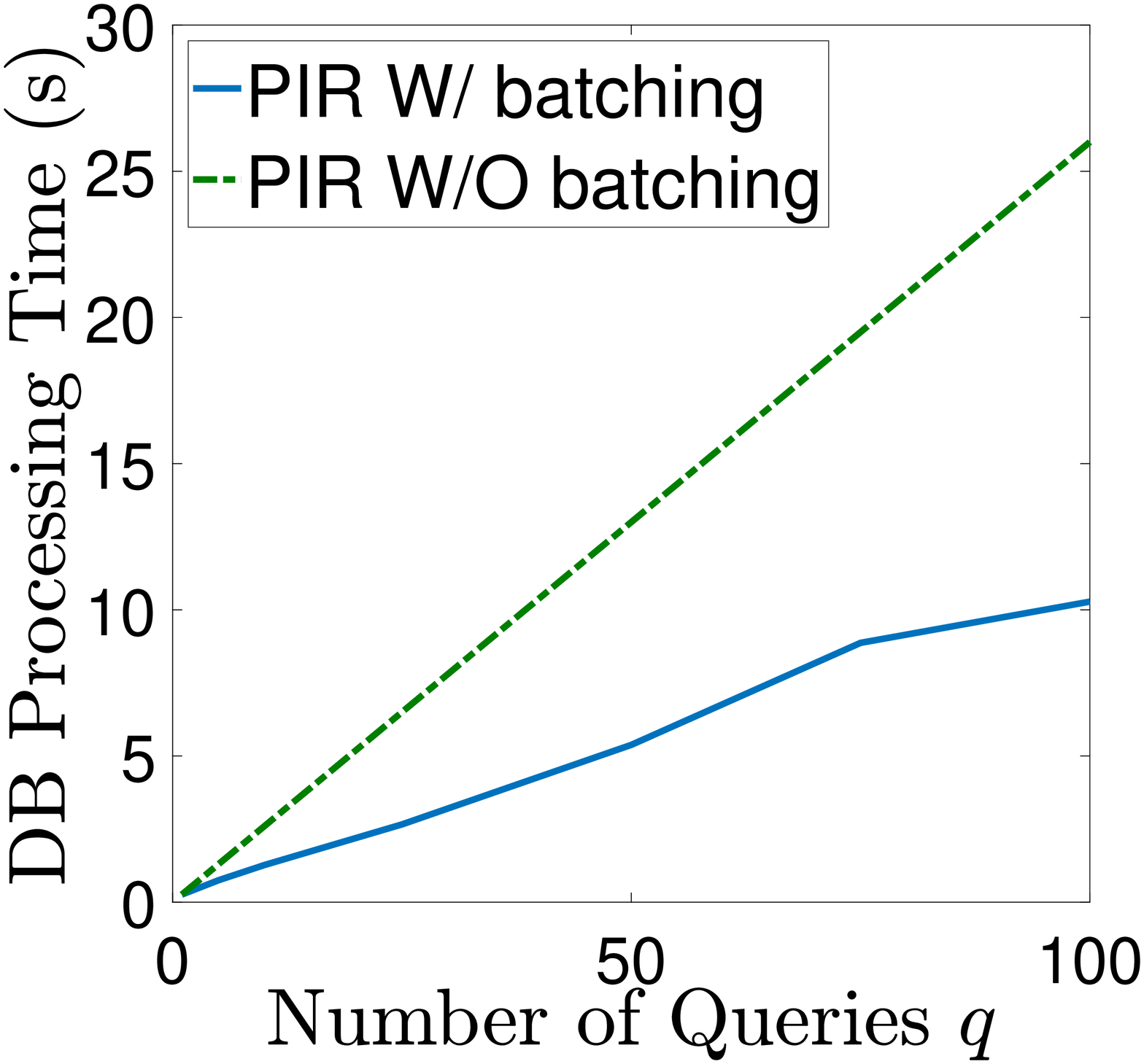}\label{fig:dbpir}}
\hfil
\subfloat[\su~\pir~delay.]{\includegraphics[width=0.162\textwidth]{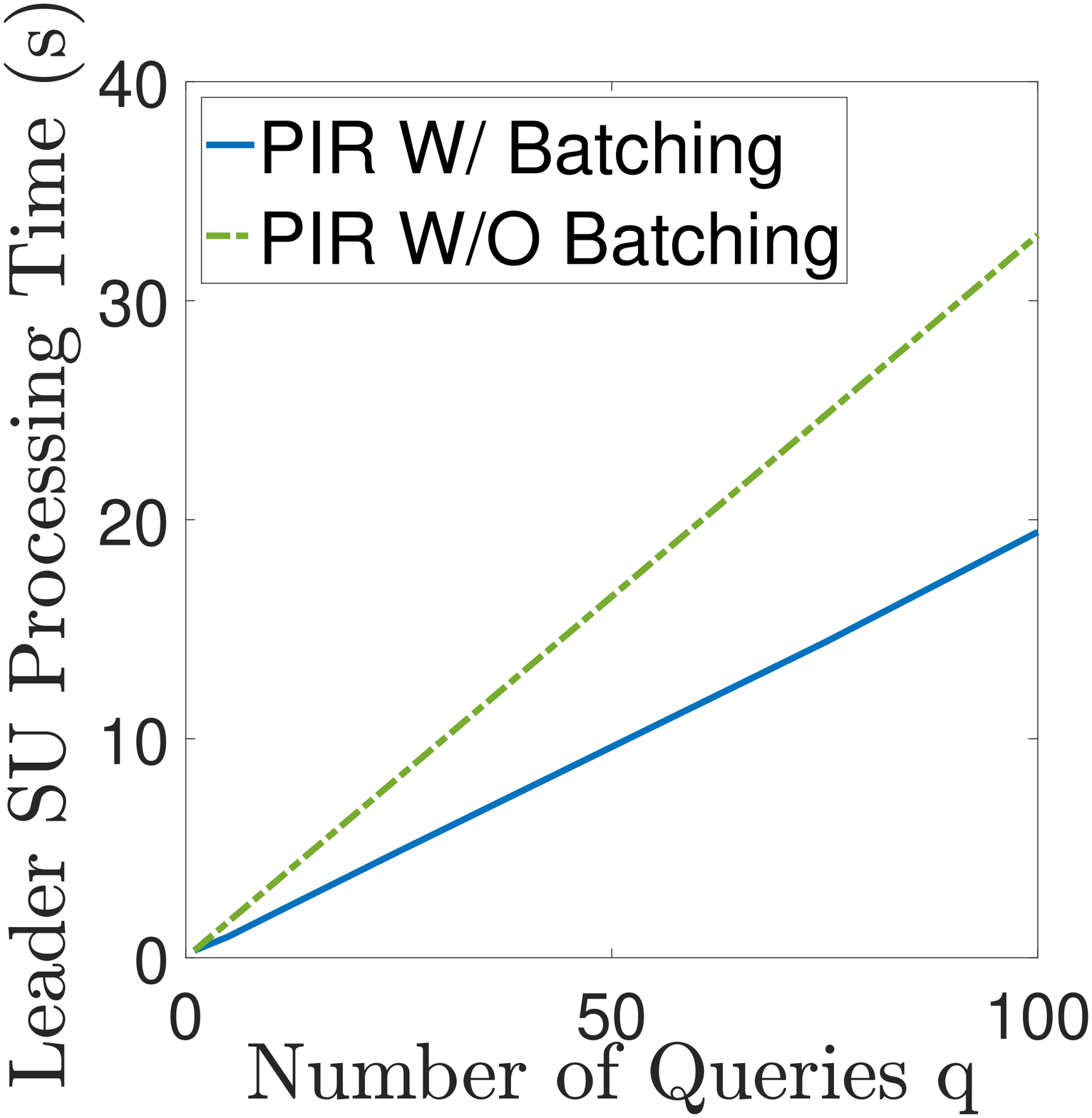}\label{fig:supir}}
\hfil
\subfloat[GoSig BFT delay.]{\includegraphics[width=0.162\textwidth]{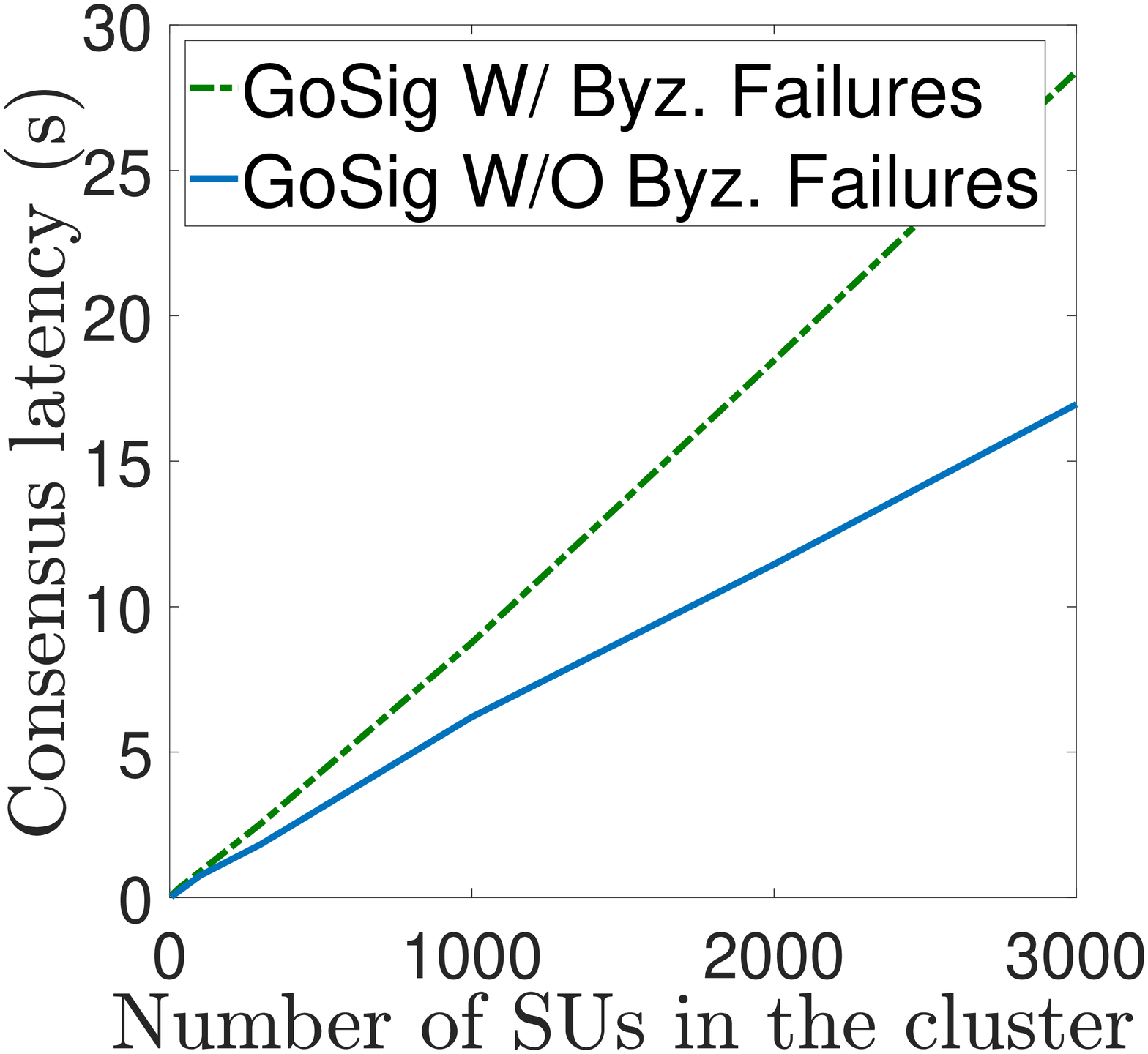}\label{fig:bft}}
\caption{Overhead of \pir~and GoSig}
\label{fig_sim}
\end{figure}

We use our Geni testbed to evaluate \protocol's multi-server PIR, BatchPIR.
As the obtained results in Figs.~\ref{fig:dbpir} and \ref{fig:supir} show, the support of query batching by BatchPIR, which allows multiple blocks to be retrieved simultaneously, reduces the overhead at both \db s' and cluster leaders' sides. We summarize the obtained results and the analytic estimation of the overhead in Table~\ref{tab:pir}.

\begin{table}[h!]
\vspace{-0.1in}
\centering  \caption{ Multi-server \pir~Overhead} \label{tab:pir}
\vspace{-2mm}
\renewcommand{\arraystretch}{1.1}{

\begin{tabular}{lll}

\toprule
{\textbf{Operation}}  & {\textbf{Analytical Cost}} &  {\textbf{Empirical Cost}}  \\
\midrule

 Leader \su's query & $q\cdot\mathcal{O}(\ell^2 r)\cdot add_\mathbb{F}$ & $4.86\;s$ \\

 \db~processing &  $q^{0.8}\cdot(\frac{8}{3} add_\mathbb{F} + mul_\mathbb{F})\cdot rs$ & $2.66\;s$ \\
 Communication &  $q\cdot(r+s)$ & $25\; MB$ \\
\bottomrule
\end{tabular}}
\flushleft{ \scriptsize{\textbf{Variables:} $\ell=7$: number of \db s, $q=25$: number of batched \pir~queries. \db~size is $\dbsize=560\; MB$, $s$: number of field $\mathbb{F}$ elements per row, $add_\mathbb{F}$ and $mul_\mathbb{F}$ denote the cost of an $\mathbb{F}$ addition and an $\mathbb{F}$ multiplication. In a field $\mathbb{F}$ of characteristic $2$, additions are equivalent to XOR and multiplications are equivalent to AND.}}
\end{table}

\begin{table*}
\vspace{-0.1in}
\centering  \caption{End-to-end Delay of \protocol~Algorithms} \label{tab:endtoend}
\vspace{-2mm}
\renewcommand{\arraystretch}{1.2}{

\begin{tabular}{lp{11cm}l}
\toprule
{\textbf{Algorithm}}  & {\textbf{Major Operations}} &  {\textbf{Total Cost}}  \\
\midrule

Alg.~\ref{alg:rekeying} Rekeying within $\cluster^{(i)}$ & $\textsc{DKG} $\:+\:$ \tgdh.\textsc{SignShareGen} $\:+\:$ \epid.\textsc{Sign} $\:+\:$\:$BFT($n_i$) &  $77.47\:s$\\

Alg.~\ref{alg:join}: Join $\cluster^{(i)}$ & $\textsc{TwoWayEPID} $\:+\:$ \textsc{Rekeying}$ & $78.12\:s$ \\
Alg.~\ref{alg:pir}: Private Spec. Query &  $\epid.\textsc{Sign} $\:+\:$ \tau\:\epid.\textsc{Verify} $\:+\:$ \textsc{BatchPIR} $\:+\:$ \:$BFT($n_i$) & $13.15\:s$ \\
Alg.~\ref{alg:usage}: Spec. Usage Notifica. & $t\:(\tgdh.\textsc{SignShareGen} $+$ \tgdh.\textsc{SignShareVerif}) $+$ \tgdh.\textsc{SignReconstruct} $+$\textsc{BFT}(\ell $+$ n_c$) & $1.85\:s$\\

\bottomrule

\end{tabular}}
\flushleft{ \scriptsize{\textbf{Parameters:} $n_i = 1000$, $t=n_i/2$, $n_c=50$ $\ell = 7$, $\tau=10$, bandwidth = $10Mbps$, $\dbsize = 560MB$, $r = 10^6$. BFT($x$): one round of BFT among $x$ parties.
}}
\vspace{-4mm}
\end{table*}

\vspace{-2mm}
\subsubsection{BFT Consensus}


Table~\ref{tab:bft} shows that the communication overhead of BFT expressed in terms of number of messages sent every consensus round is quasi-linear in the size of the cluster, $n_i$, which translates into a total communication overhead of $\mathcal{O}(n_i^2\log n_i)$.
%
%
In this experiment, we also set the throughput between the nodes to $10\;Mbps$ and the propagation delay among \su s to $20\; ms$ and simulate the protocol to estimate the time it takes to reach a consensus over a block. Our results, depicted in Fig.~\ref{fig:bft}, show that even for a cluster of size as large as 1000 \su s, a consensus is reachable in less than $7 \;s$ even if up to $1/3$ of the \su s are Byzantine. The overhead of BFT depends heavily on the number of participants and the number of signature verifications required by each participant. Therefore, BFT will have a different cost for each of \protocol's algorithms. For instance in \textsc{Rekeying}, BFT will take as long as $76\:s$ since each \su~will need to verify the signatures of all other \su s in $\cluster^{(i)}$ included in the block submitted by the leader at step~7 of Alg.~\ref{alg:rekeying}.


\begin{table}
\vspace{-0.1in}
\centering  \caption{ BFT complexity} \label{tab:bft}
\vspace{-2mm}
\renewcommand{\arraystretch}{1.1}{

\begin{tabular}{lll}
\toprule
{\textbf{Operation}}  & {\textbf{Analytical Cost}} &  {\textbf{Empirical Cost}}  \\
\midrule

Communication per user & $\mathcal{O}(n_i\log n_i)$ &  $\propto 3000$ messages\\

Consensus w/o failures & $\mathcal{O}(n_i^2\log n_i)$ & $4.3\;s$ \\
Consensus w/ failures & $\mathcal{O}(n_i^2\log n_i)$ & $6.32\;s$ \\
\bottomrule

\end{tabular}}
\flushleft{ \scriptsize{\textbf{Parameters:} $n_i = 1000$, bandwidth = $10Mbps$, $1$ signature verification per \su.
}}
\end{table}

\subsubsection{End-to-end Delay}
We provide in Table~\ref{tab:endtoend} the end-to-end delays caused by \protocol's different algorithms, ignoring Byzantine faultiness for simplicity.
%
%
%
Observe that the \textsc{Rekeying} has the highest cost, which is invoked mainly when a membership change occurs. One way to address this is by setting \textsc{Rekeying} frequency small, and have joining \su s wait a little longer before they join the system.
Another way to further reduce the cost in most of these algorithms is by using different quorums of users every BFT round. This will reduce the overhead but will also impact the security guarantees and robustness against failures. Despite the relatively high cost of these algorithms, note that these operations are expected to be invoked only every few hours, as it is the case for TVWS, which requires \su s to query \db s every 24 hours.


\section{Conclusion}
\label{sec:conclusion}
We propose \protocol, a trustworthy framework for \sas~that preserves \su s' operational privacy while adhering to regulatory requirements mandated by FCC in the 3.5 GHz CBRS band. \protocol~achieves this by synergizing state-of-the-art cryptographic mechanisms with the blockchain technology. We show the privacy benefits of \protocol~through security analysis, simulation and experimentation.


\section*{Acknowledgment}
This work was supported in part by the US National Science Foundation under NSF awards CNS-1162296 and CNS-1652389.
\vspace{-4mm}
\small{
\bibliographystyle{IEEEtran}
\bibliography{IEEEabrv,references}

\begin{thebibliography}{10}
\providecommand{\url}[1]{#1}
\csname url@samestyle\endcsname
\providecommand{\newblock}{\relax}
\providecommand{\bibinfo}[2]{#2}
\providecommand{\BIBentrySTDinterwordspacing}{\spaceskip=0pt\relax}
\providecommand{\BIBentryALTinterwordstretchfactor}{4}
\providecommand{\BIBentryALTinterwordspacing}{\spaceskip=\fontdimen2\font plus
\BIBentryALTinterwordstretchfactor\fontdimen3\font minus
  \fontdimen4\font\relax}
\providecommand{\BIBforeignlanguage}[2]{{%
\expandafter\ifx\csname l@#1\endcsname\relax
\typeout{** WARNING: IEEEtran.bst: No hyphenation pattern has been}%
\typeout{** loaded for the language `#1'. Using the pattern for}%
\typeout{** the default language instead.}%
\else
\language=\csname l@#1\endcsname
\fi
#2}}
\providecommand{\BIBdecl}{\relax}
\BIBdecl

\bibitem{federal2015report}
FCC, ``Report and order and second further notice of proposed rulemaking,
  number 15-47, gn docket no. 12-354. {FCC},'' April 2015.

\bibitem{federal2016report}
------, ``Order on reconsideration and second report and order, number 16-55,
  gn docket no. 12-354. {FCC},'' May 2016.

\bibitem{ye2016overview}
Y.~Ye, D.~Wu, Z.~Shu, and Y.~Qian, ``Overview of lte spectrum sharing
  technologies,'' \emph{IEEE Access}, vol.~4, pp. 8105--8115, 2016.

\bibitem{chen2015protocol}
V.~Chen, S.~Das, L.~Zhu, J.~Malyar, and P.~McCann, ``Protocol to access
  white-space (paws) databases,'' Tech. Rep., 2015.

\bibitem{clark2018trading}
M.~A. Clark and K.~Psounis, ``Trading utility for privacy in shared spectrum
  access systems,'' \emph{IEEE/ACM Transactions on Networking (TON)}, vol.~26,
  no.~1, pp. 259--273, 2018.

\bibitem{marshall2017three}
P.~Marshall, \emph{Three-tier Shared Spectrum, Shared Infrastructure, and a
  Path to 5G}.\hskip 1em plus 0.5em minus 0.4em\relax Cambridge University
  Press, 2017.

\bibitem{winnf2017cbrscomm}
W.~I. Forum, ``Cbrs communications security technical specification,
  winnf-15-s-0065,'' April 2017.

\bibitem{winnf2016cbrsthreat}
------, ``Cbrs threat model technical report, winnf-15-p-0089,'' May 2016.

\bibitem{grissa2017locationsurvey}
M.~Grissa, B.~Hamdaoui, and A.~A. Yavuza, ``Location privacy in cognitive radio
  networks: A survey,'' \emph{IEEE Communications Surveys \& Tutorials},
  vol.~19, no.~3, pp. 1726--1760, 2017.

\bibitem{khalfi2018airmap}
B.~Khalfi, B.~Hamdaoui, and M.~Guizani, ``Airmap: Scalable spectrum occupancy
  recovery using local low-rank matrix approximation,'' in \emph{Global
  Communications Conference (GLOBECOM), 2018 IEEE}.

\bibitem{grissa2018unleashing}
M.~Grissa, B.~Hamdaoui, and A.~A. Yavuz, ``Unleashing the power of multi-server
  pir for enabling private access to spectrum databases,'' \emph{IEEE
  Communications Magazine}, vol.~56, pp. 171--177, December 2018.

\bibitem{chor1995private}
B.~Chor, O.~Goldreich, E.~Kushilevitz, and M.~Sudan, ``Private information
  retrieval,'' in \emph{Foundations of Computer Science, 1995. Proceedings.,
  36th Annual Symposium on}.\hskip 1em plus 0.5em minus 0.4em\relax IEEE, 1995,
  pp. 41--50.

\bibitem{vukolic2017rethinking}
M.~Vukoli{\'c}, ``Rethinking permissioned blockchains,'' in \emph{Proceedings
  of ACM Workshop on Blockchain, Cryptocurrencies and Contracts}, 2017.

\bibitem{brickell2012enhanced}
E.~Brickell and J.~Li, ``Enhanced privacy id: A direct anonymous attestation
  scheme with enhanced revocation capabilities,'' \emph{IEEE Transactions on
  Dependable and Secure Computing}, vol.~9, no.~3, pp. 345--360, 2012.

\bibitem{rackoff1991non}
C.~Rackoff and D.~R. Simon, ``Non-interactive zero-knowledge proof of knowledge
  and chosen ciphertext attack,'' in \emph{Annual International Cryptology
  Conference}.\hskip 1em plus 0.5em minus 0.4em\relax Springer, 1991, pp.
  433--444.

\bibitem{boldyreva2003threshold}
A.~Boldyreva, ``Threshold signatures, multisignatures and blind signatures
  based on the gap-diffie-hellman-group signature scheme,'' in \emph{Int'l
  Workshop on Public Key Cryptography}.\hskip 1em plus 0.5em minus 0.4em\relax
  Springer, 2003, pp. 31--46.

\bibitem{gennaro1999secure}
R.~Gennaro, S.~Jarecki, H.~Krawczyk, and T.~Rabin, ``Secure distributed key
  generation for discrete-log based cryptosystems,'' in \emph{Int'l Conf. on
  the Theory and App. of Crypto Tech.}\hskip 1em plus 0.5em minus 0.4em\relax
  Springer, 1999, pp. 295--310.

\bibitem{shamir1979share}
A.~Shamir, ``How to share a secret,'' \emph{Communications of the ACM},
  vol.~22, no.~11, pp. 612--613, 1979.

\bibitem{goldfeder2017escrow}
S.~Goldfeder, J.~Bonneau, R.~Gennaro, and A.~Narayanan, ``Escrow protocols for
  cryptocurrencies: How to buy physical goods using bitcoin,'' in \emph{Int'l
  Conf. on Financial Crypto and Data Security}.\hskip 1em plus 0.5em minus
  0.4em\relax Springer, 2017.

\bibitem{chen2007cogmesh}
T.~Chen, H.~Zhang, G.~M. Maggio, and I.~Chlamtac, ``Cogmesh: A cluster-based
  cognitive radio network,'' in \emph{2007 2nd IEEE International Symposium on
  New Frontiers in Dynamic Spectrum Access Networks}.

\bibitem{lueks2015sublinear}
W.~Lueks and I.~Goldberg, ``Sublinear scaling for multi-client private
  information retrieval,'' in \emph{International Conference on Financial
  Cryptography and Data Security}.\hskip 1em plus 0.5em minus 0.4em\relax
  Springer, 2015, pp. 168--186.

\bibitem{castro1999practical}
M.~Castro, B.~Liskov \emph{et~al.}, ``Practical byzantine fault tolerance,'' in
  \emph{OSDI}, vol.~99, 1999, pp. 173--186.

\bibitem{miracl}
M.~Integer and C.~Rational~Arithmetic, ``C++ library (miracl),''
  \url{https://github.com/miracl/MIRACL}, 2013, accessed: 2018-06-02.

\bibitem{geni}
``Global environment for network innovations,'' \url{https://www.geni.net/ }.

\bibitem{percy}
``Percy++ library,'' \url{http://percy.sourceforge.net}, accessed: 2018-06-14.

\bibitem{tblsdfinity}
``Threshold bls dfinity implementation,''
  \url{https://github.com/dfinity/random-beacon }, accessed: 2018-06-02.

\bibitem{epidsdk}
``The intel(r) enhanced privacy id software development kit,''
  \url{https://github.com/Intel-EPID-SDK}, accessed: 2018-06-02.

\bibitem{keylength}
\url{https://www.keylength.com/}, accessed: 2018-06-02.

\end{thebibliography}
}

\end{document}